\documentclass[review]{elsarticle}
\usepackage{lineno,hyperref,threeparttable,amssymb,aas_macros,appendix}
\biboptions{authoryear}

\begin{document}

\begin{frontmatter} 

\title{Jetting during oblique impacts of spherical impactors}

\author[a]{Shigeru Wakita\corref{corresponding}}
\cortext[corresponding]{Corresponding author}
\author[a]{Brandon Johnson}
\author[a]{C. Adeene Denton}
\author[b]{Thomas M. Davison}
\address[a]{Department of Earth, Atmospheric, and Planetary Sciences, Purdue University, West Lafayette, IN, 47907, USA}
\address[b]{Department of Earth Science and Engineering, Imperial College London, London, UK}

\begin{abstract}
During the early stages of an impact a small amount material may be jetted and ejected at speeds exceeding the impact velocity. 
Jetting is an important process for producing melt during relatively low velocity impacts. 
How impact angle affects the jetting process has yet to be fully understood.
Here, we simulate jetting during oblique impacts using the iSALE shock physics code. 
Assuming both the target and impactor have the same composition (dunite), we examine the jetted material which exceeds the impact velocity.
Our results show that oblique impacts always produce more jetted ejecta than vertical impacts, 
except for grazing impacts with impact angles $<$ 15$^\circ$. 
A 45$^\circ$ impact with an impact velocity of 3 km/s produces jetted material equal to $\sim$ 7\% of the impactor mass. 
This is 6 times the jetted mass produced by a vertical impact with similar impact conditions. 
We also find that the origin of jetted ejecta depends on impact angle; for impact angles less than 45$^\circ$, 
most of the jet is composed of impactor material, while at higher impact angles the jet is dominated by target material. 
Our findings are consistent with previous experimental work.
In all cases, jetted materials are preferentially distributed downrange of the impactor.  
\end{abstract}


\end{frontmatter} 

\section{Introduction} \label{sec:int}
Impact jetting occurs during the earliest stage of impact cratering when the impactor is still making contact with the target \citep[e.g.,][]{Birkhoff:1948aa,Walsh:1953aa}.
The jet contains the first material ejected, is highly shocked, and is ejected at speed exceeding the impact velocity. 
The total mass of jetted ejecta is small, typically less than a few percent of the impactor mass. 
Impact jetting has been proposed as a mechanism 
for producing tektites \citep{Vickery:1993aa,Stoffler:2013aa},
producing melt in lightly shocked material during grain-on-grain collision \citep{Kieffer:1977aa},
and producing chondrules \citep{Kieffer:1975aa,Johnson:2015aa,Wakita:2017aa}.
Jetting has also been proposed as a critical component of vaporization during massive impacts \citep{Melosh:1986aa}, 
and to affect the final ice-rock-ratio of icy bodies \citep{McKinnon:1989aa,McKinnon:1989ab}. 
However, jetting, especially during oblique impacts, remains poorly understood.

Experimental and theoretical studies have explored jetting during planetary impacts using thin plate theory
\citep[e.g.,][]{Kieffer:1975aa,Melosh:1986aa,Vickery:1993aa}.
Thin plate theory describes jetting during the symmetric collision of two thin plates 
and can predict the jet velocity and the critical angle above which jetting occurs as a function of plate material and impact velocity
\citep{Birkhoff:1948aa,Walsh:1953aa}. 
However, thin plate theory fails when applied to planetary impacts, which are better described by the impact between a spherical impactor and a plate than of two thin plates,
including predictions of ejecta temperature and the ratio of target to impactor material in jets produced by spherical impactors \citep{Sugita:1999aa}.
Numerical models \citep{Johnson:2014aa} and laboratory experiments \citep{Kurosawa:2015aa} show 
that the maximum velocities of jets produced during the impact of spherical bodies are about half the velocity predicted by thin plate theory.  

\citet{Vickery:1993aa} estimated that very oblique impacts could have up to ten times more jetted material than vertical impacts with the same impact velocity. 
While these estimates are based on thin plate theory,
which does not accurately describe the jetting physics during impact of spherical bodies \citep{Yang:1995aa, Kurosawa:2015aa},
oblique impacts should produce more jetted material than head-on collisions 
as experimental results of the spherical impactor indicate \citep{Sugita:1999aa}. 
However, without direct simulation in three dimensions, 
we cannot quantify how impact angle will affect the amount of jetted material produced by a given impact.

Most impacts are believed to be oblique; a typical impact angle is expected to be 45$^\circ$ \citep{Shoemaker:1962aa}.
The effect of impact angle on crater formation has been studied experimentally \citep[e.g.,][]{Gault:1978aa,Schultz:1990aa,Burchell:1998aa} 
and numerically \citep[e.g.,][]{Davison:2011aa,Elbeshausen:2009aa}.
Highly oblique impacts can produce elongated craters as found on Mars, Venus, and the Moon \citep{Bottke:2000aa}. 
Even moderately oblique impacts have a significant effect on distal impact ejecta, producing a wedge of avoidance in the uprange direction \citep{Gault:1978aa},
and elongated craters produced by grazing impacts exhibit in a butterfly-like distal ejecta pattern \citep{Gault:1978aa,Schultz:1990aa,Speyerer:2016aa}.
How impact angle affects the distribution of the very fastest and earliest ejecta remains uncertain.
Here, we simulate oblique impacts to determine how impact angle affects the jetting process. 
This includes determining how the amount and distribution of jetted material depends on impact angle.
In Section \ref{sec:met}, we describe the setup for our numerical simulations.
We then present results from our simulations (Section \ref{sec:res}) 
and discuss the implications for our understanding of the unique physics surrounding jetted material during impact (Section \ref{sec:dis}).

\section{Methods}\label{sec:met}

We simulate jetting during oblique impacts using the iSALE-3D shock physics code,
which includes both a strength model and the capability to simulate porous compaction 
\citep{Hirt:1974aa,Collins:2004aa,Melosh:1992aa,Ivanov:1997aa,Wunnemann:2006aa,Collins:2011aa}.
We assume that our spherical impactor and flat target have the same dunite composition \citep{Benz:1989aa}, 
as used in previous jetting work \citep[e.g.,][]{Johnson:2014aa}.
We note these results are only strictly valid for a dunite impactor and target. 
More compressible materials are known to have a lower jetting efficiency \citep{Johnson:2014aa}. 
Jetting during the impact of a projectile with distinct composition from the target will also affect jetting dynamics \citep{Sugita:1999aa}. 
We expect, however, that the downrange momentum of the impactor in oblique impacts will still tend to enhance jetting regardless of composition.
The material properties for dunite used in iSALE-3D are described in Table \ref{tab:isale}.

In our simulations, we vary impact angle ($\theta_{\rm imp}$) and impact velocity ($v_{\rm imp}$).
Impact angle in our simulations is measured from the horizontal surface of the target, 
such that $\theta_{\rm imp}=90^\circ$ is a vertical impact. 
We simulate a range of impact angles from 15$^\circ$ to 90$^\circ$ in 15$^\circ$ increments, 
with impactor velocities of 2 km/s, 3 km/s, and 5 km/s. 
This range of velocities spans the expected range of impact velocities during planetesimal accretion, 
which depends on the size of the target bodies \citep[i.e., the escape velocity of the system,][]{Kokubo:2000aa,Johnson:2015aa,Hasegawa:2016aa}.
This is also the velocity range where the jetted mass is highest for vertical impacts of dunite impactors on dunite targets
\citep{Johnson:2015aa, Wakita:2017aa}.

We record the behavior of jetted ejecta in iSALE-3D 
using Lagrangian tracer particles to track material position, velocity, pressure, and temperature. 
Note that the tracer particles, which track the motion of a parcel of material, are initially placed at the center of each cell and 
we calculate their mass using their initial spacing, locations, and density \citep[e.g.,][]{Johnson:2014ab}.
We focus on tracer particles since they allow us to determine material provenance as well as mass.
High resolution is required to accurately evaluate the velocities, ejection angles, and mass of jetted ejecta.
We use a fixed impactor diameter of $D_{\rm imp}$ = 1 km,
with a spatial resolution of 5 m or 100 cells per projectile radius.
Previous work showed that for 10-km-diameter bodies, jetting occurs in a near-surface zone at most $\sim$ 400 m deep \citep{Wakita:2017aa}.
As such, the high-resolution zone of our model incorporates the lower half of the impactor and a thin surface layer of target material.
We use this setting for most of our runs. 
iSALE-3D divides simulations into a high-resolution zone and an extension zone;
in the extension zone cell size increases with distance from the high-resolution zone. 
To avoid errors introduced by motion through the extension zone, 
we only track tracer particles in the high resolution zone. 
To confirm that our high-resolution zone appropriately resolves the jetting process, 
we compare our iSALE-3D results of vertical impacts with iSALE-2D runs using the same resolution (\ref{sec:app}).
We also ensure that all jetted material originates from and is tracked within the high resolution zone. 
We confirm that a high resolution zone encompassing half of the impactor resolves jetted material well for most oblique impact scenarios.
For some oblique impact scenarios ($\theta_{\rm imp} = 30^\circ$ at $v_{\rm imp}$ = 5 km/s and $\theta_{\rm imp} = 45^\circ$ at $v_{\rm imp}$ = 2, 3, and 5 km/s),
we enlarge the high-resolution zone in the vertical direction to include all jetted impactor material.

\begin{threeparttable}
\caption{iSALE input parameters}
\begin{tabular}{ll}
\hline
\hline
Description & Values \\
\hline
Equation of state &  ANEOS \\
Bulk material of impactor/target & dunite \tnote{a}\\
Solidus temperature & 1373 K \tnote{b} \\
Simon approximation constant A & 1520 MPa \tnote{c} \\
Simon approximation exponent C & 4.05 \tnote{c} \\
Poisson's ratio & 0.25 \tnote{d} \\
Thermal softening parameter & 1.1 \tnote{d} \\
Strength model & Rock \tnote{e	} \\
Cohesion (damaged) & 0.01 MPa \tnote{d} \\
Cohesion (undamaged) & 10 MPa \tnote{d} \\
Frictional coefficient (damaged) & 0.6 \tnote{d} \\
Frictional coefficient (undamaged) & 1.2 \tnote{d} \\
Strength at infinite pressure & 3.5 GPa \tnote{d} \\
Damage model & Ivanov \tnote{e} \\
Minimum failure strain &  $10^{-4}$ \tnote{d} \\
Damage model constant &  $10^{-11}$ \tnote{d} \\
Threshold pressure for damage model &  300 MPa \tnote{d} \\
\hline
\end{tabular}
\begin{tablenotes}
\item[a] \citet{Benz:1989aa}
\item[b] \citet{Davison:2016aa}
\item[c] \citet{Davison:2010aa}
\item[d] \citet{Kurosawa:2018aa}
\item[e] \citet{Collins:2004aa}
\end{tablenotes}
\label{tab:isale}
\end{threeparttable}

\section{Results} \label{sec:res}

\begin{figure}
\includegraphics[clip,width=\textwidth]{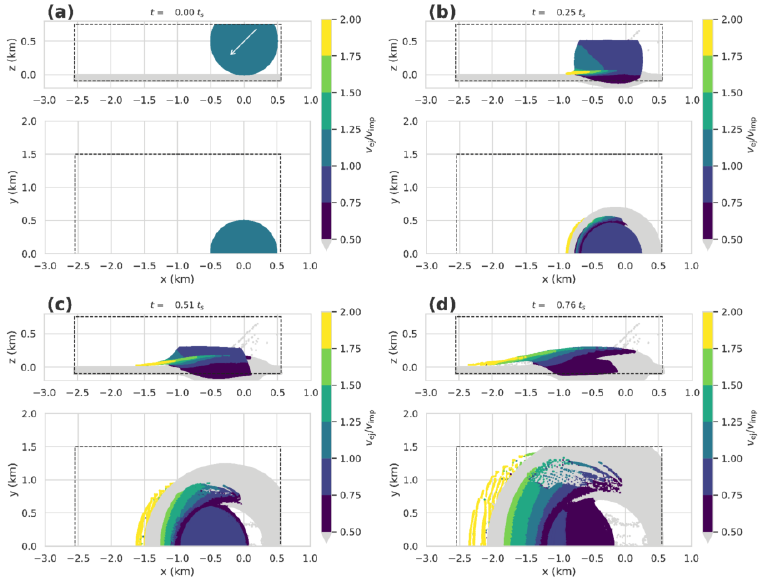}
\caption{
Time series for our fiducial model ($v_{\rm imp}$ = 3 km/s and $\theta_{\rm imp}$ = 45$^\circ$).
Color represents tracer particle velocity ($v_{\rm ej}$) normalized by the impact velocity.
The top panels are a cross-sectional view,
while the bottom panels are a top view on the impact from above.
Each panel represents (a) $t$ = 0.0 $t_s$, (b) $t$ = 0.25 $t_s$, (c) $t$ = 0.51 $t_s$, and (d) 0.76 $t_s$ after the impact, respectively.
White arrow in top-left panel shows the direction of impactor.
Note the impactor is spherical, but only contains tracer particles in lower part,
due to our setting of high resolution zone (shown as dashed line).
\label{fig:45totalv}}
\end{figure}

\subsection{Effect of impact angle on distribution of jetted material} \label{sec:res-angle}
\begin{figure}
\includegraphics[clip,width=\textwidth]{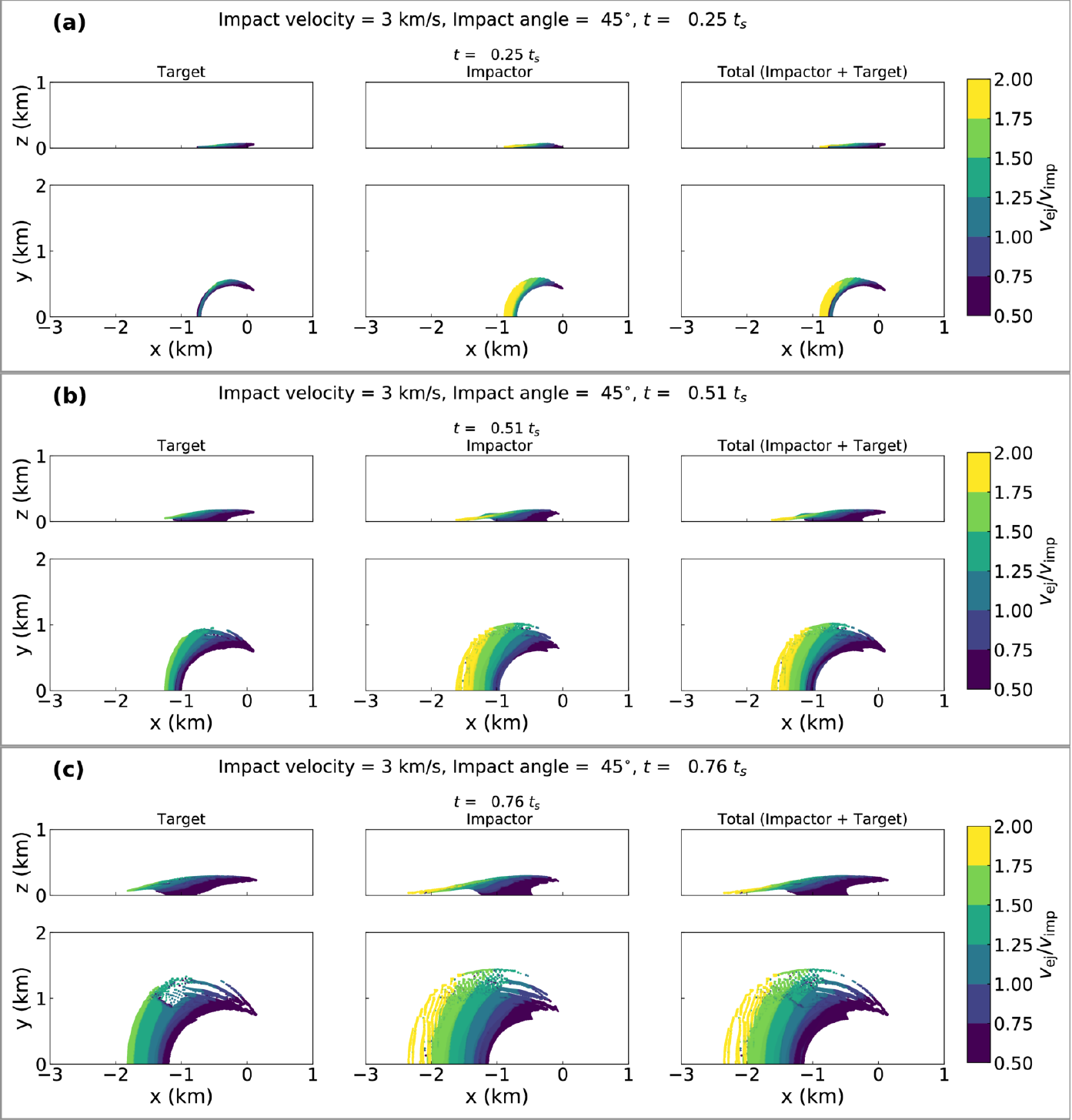}
\caption{
Material ejected at (a) $t$ = 0.25 $t_s$, (b) $t$ = 0.51 $t_s$, and (c) $t$ = 0.76 $t_s$ 
for our fiducial model ($v_{\rm imp}$ = 3 km/s and $\theta_{\rm imp}$ = 45$^\circ$).
Color represents ejecta velocity ($v_{\rm ej}$) normalized by the impact velocity.
Top panels are a cross-sectional view, while the bottom panels are a bird's eye view on the impact from above.
Ejecta produced by the target are shown on the left, 
ejecta from the impactor in the middle, 
and the total ejecta produced from the impact on the right, respectively.
\label{fig:45}}
\end{figure}

Figure \ref{fig:45totalv} shows a time series of our fiducial model, 
an oblique impact with $\theta_{\rm imp}$ = 45$^\circ$ and $v_{\rm imp}$ = 3 km/s.
Each panel represents $t$ = 0.0 $t_s$, 0.25 $t_s$, 0.51 $t_s$, and 0.76 $t_s$ after the impact, respectively,
where $t_s$ is the characteristic time for contact and compression defined by $t_s = D_{\rm imp}/(v_{\rm imp} \sin(\theta_{\rm imp}))$.
The tracer particles shown in Figure \ref{fig:45totalv} reflect our setup of high-resolution zone, 
which includes the lower portion of the spherical impactor and the near-surface portion of the target.
Figure \ref{fig:45} illustrates only the tracer particles whose velocity is higher than the half of the impact velocity
with a positive velocity in vertical direction. 
Faster ejecta ($v_{\rm ej}/v_{\rm imp} > 1$) is focused in the downrange direction and 
ejected at very low angles as evidenced by its nearly horizontal motion.
The higher velocity ejecta ($v_{\rm ej}/v_{\rm imp} > 1.6$) are preferentially distributed downrange and originate from the impactor. 

\begin{figure}
\includegraphics[clip,width=\textwidth]{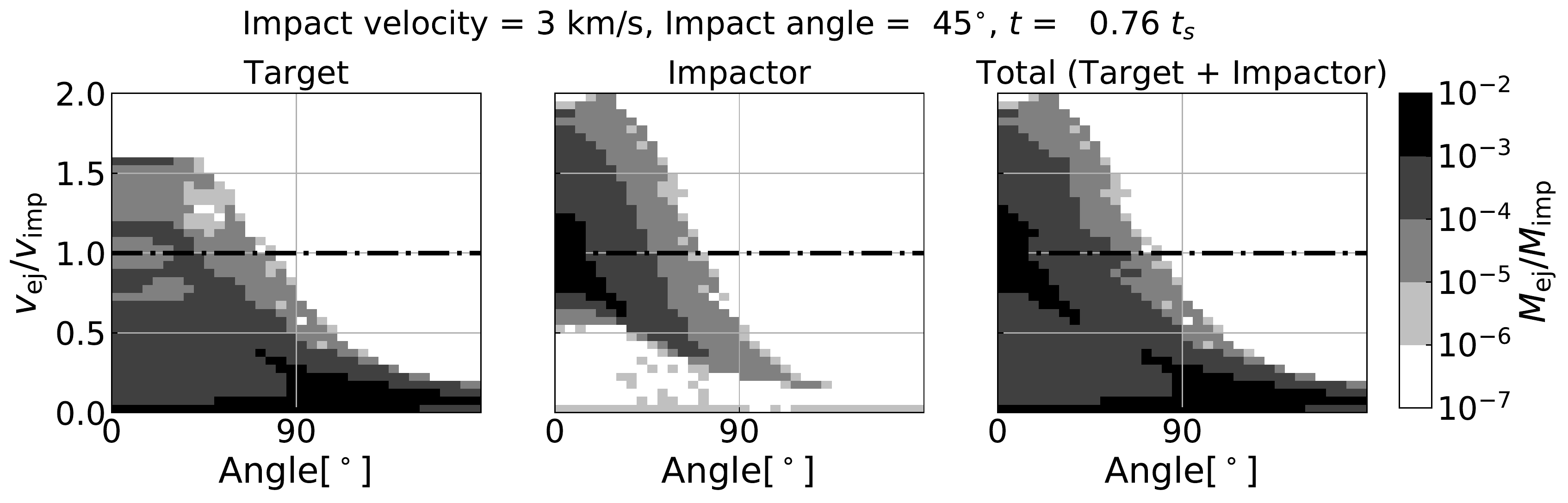}
\caption{Heatmap of ejecta density as a function of ejection angle and velocity for our fiducial model ($v_{\rm imp}$ = 3 km/s and $\theta_{\rm imp}$ = 45$^\circ$).
Ejection angle is defined with respect to the orientation of the oblique impactor (downrange is  0$^\circ$, uprange 180$^\circ$).
Note ejection angle is measured in the azimuthal direction, not in vertical direction (elevation angle).
Gray contours represent mass of ejecta $M_{\rm ej}$ normalized by mass of impactor $M_{\rm imp}$:
each bin of ejection angle and velocity represents the amount of ejecta in the bin 
(the dark-colored bin has more ejecta than the light-colored bin).
Note that angle bins are 5 degree wide and velocity bins have a width of 0.05 $v_{\rm ej}/v_{\rm imp}$.
Ejecta produced by the target are shown on the left, 
ejecta from the impactor in the middle, 
and the total ejecta produced from the impact on the right, respectively.
\label{fig:45hm}}
\end{figure}

Figure \ref{fig:45hm} illustrates the relationship between ejecta velocity 
and azimuthal ejection angle as defined with respect to the orientation of the impactor 
 (downrange is  0$^\circ$, uprange 180$^\circ$).
As Figure \ref{fig:45hm} indicates, jetted ejecta ($v_{\rm ej}/v_{\rm imp} > 1$) is largely distributed downrange,
with azimuths less than 90$^\circ$. 
Azimuthal angle shallows as the ejecta velocity increases.
Because the impactor is traveling in the downrange direction on impact,
downrange ejecta acquires higher velocity than ejecta not aligned with the impactor.
Although the distribution of ejecta produced from the target and the impactor in Figure \ref{fig:45} is similar,
the maximum velocity of ejecta produced from impactor material is $v_{\rm ej}/v_{\rm imp} = 2$,
which is higher than ejecta produced from the target of $v_{\rm ej}/v_{\rm imp} = 1.6$.

\begin{figure}
\includegraphics[clip,width=\textwidth]{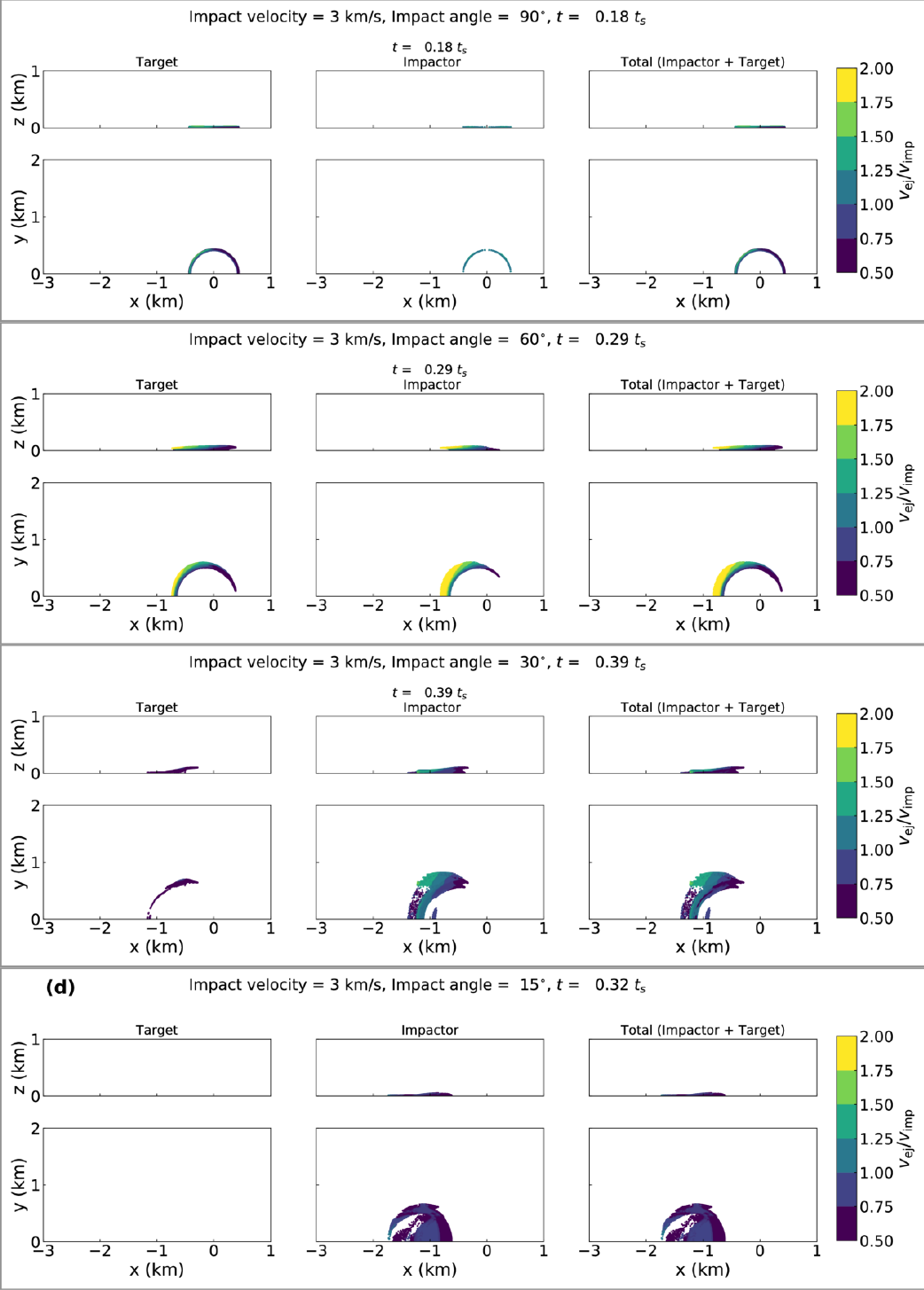}
\caption{
Behavior of ejected material for 
$\theta_{\rm imp}$ = 90$^\circ$, 60$^\circ$, 30$^\circ$, and 15$^\circ$ (top to bottom).
Same color scheme and viewing geometries as Figure \ref{fig:45}.
\label{fig:viewsh}}
\end{figure}

\begin{figure}
\includegraphics[clip,width=\textwidth]{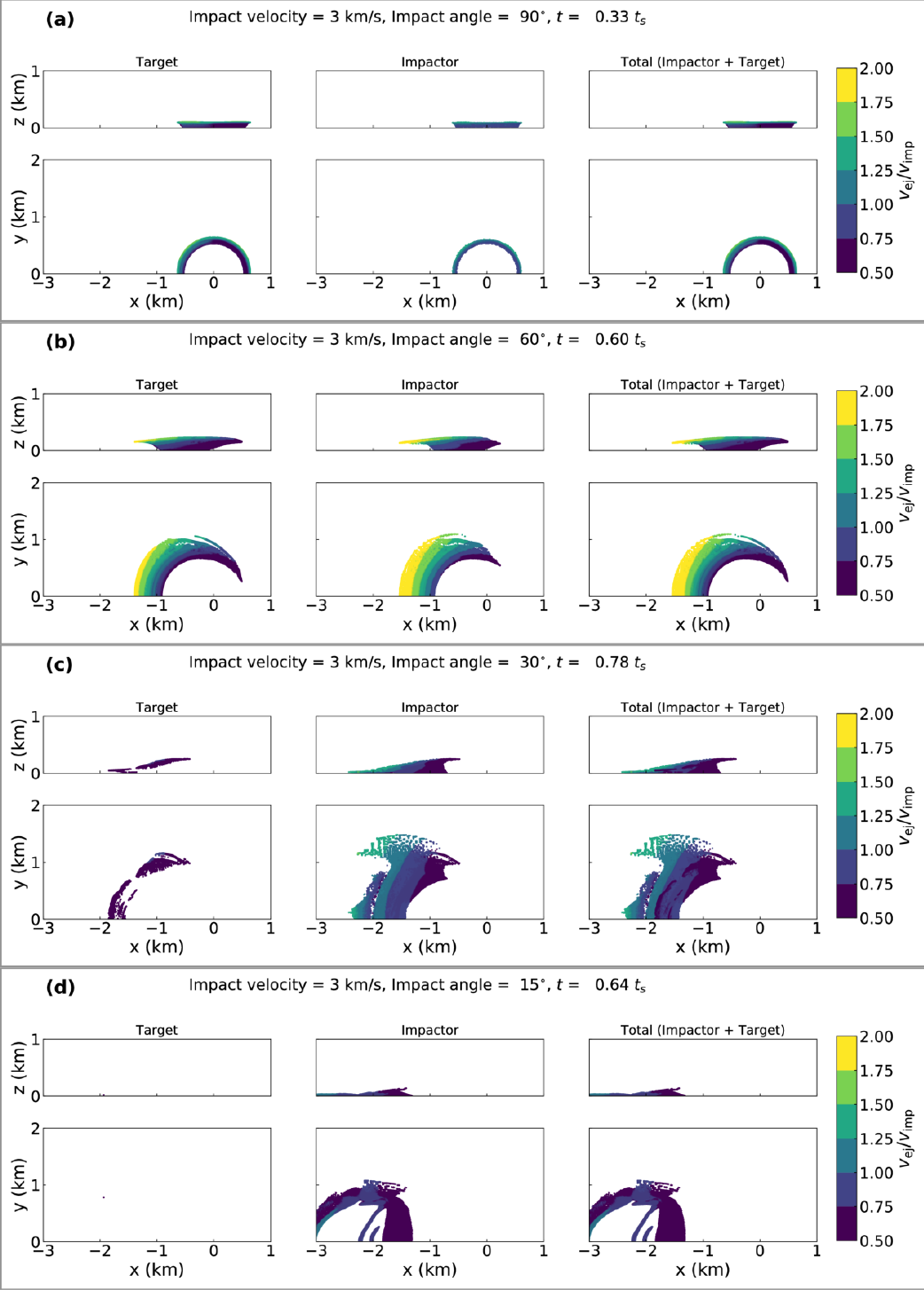}
\caption{
Same as Figure \ref{fig:viewsh}, but at later time.
\label{fig:views}}
\end{figure}

\begin{figure}
\includegraphics[clip,width=\textwidth]{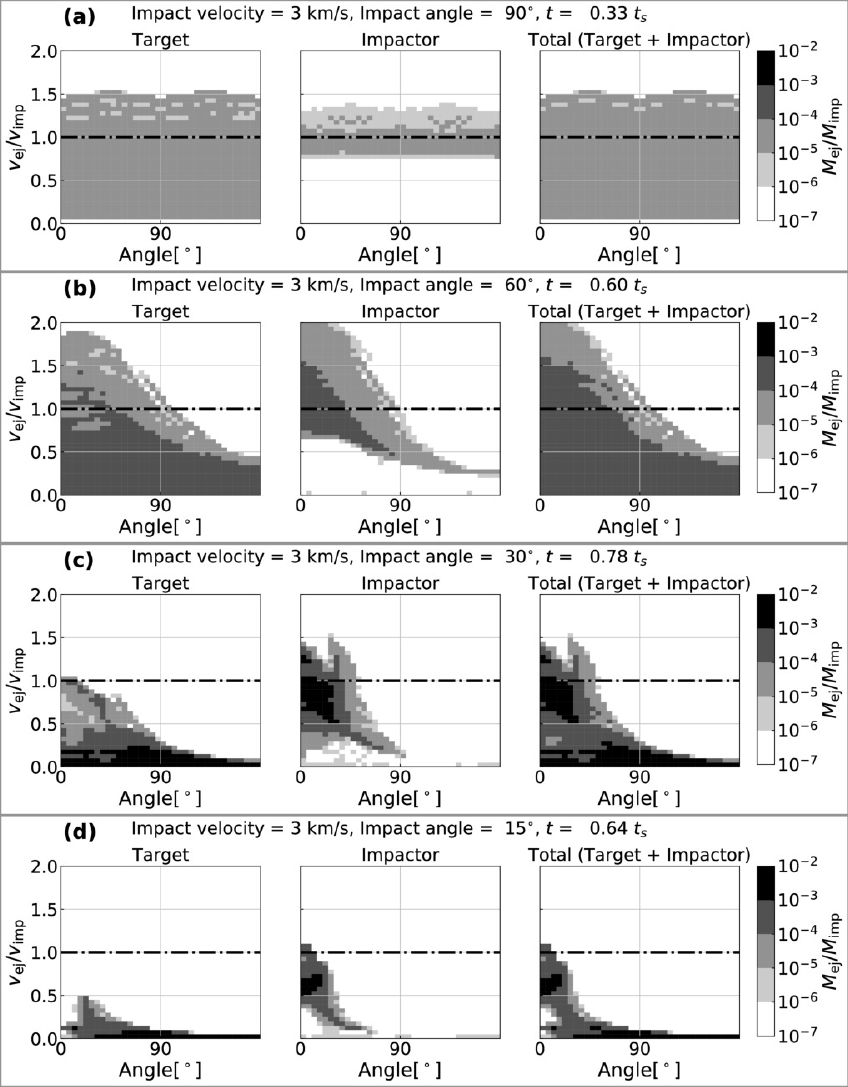}
\caption{Same color scheme and measurements as Figure \ref{fig:45hm} produced for 
$\theta_{\rm imp}$ = 90$^\circ$, 60$^\circ$, 30$^\circ$, and 15$^\circ$ (top to bottom).
\label{fig:hms}}
\end{figure}

Figures \ref{fig:viewsh}, \ref{fig:views} and \ref{fig:hms} show the resulting high-velocity ejecta trends for impact scenarios incorporating
$\theta_{\rm imp}$ = 90$^\circ$, 60$^\circ$, 30$^\circ$, and 15$^\circ$. 
For a head-on impact ($\theta_{\rm imp}$ = 90$^\circ$)
ejecta are distributed symmetrically from the impact point (defined as the origin, 0, 0) (Figures \ref{fig:viewsh}a, \ref{fig:views}a, \ref{fig:hms}a). 
Ejecta behavior for $\theta_{\rm imp}$ = 60$^\circ$
is similar to $\theta_{\rm imp}$ = 45$^\circ$, our fiducial case (Figures \ref{fig:viewsh}b, \ref{fig:views}b and \ref{fig:hms}b);
however, as impact angle continues to decrease ($\theta_{\rm imp}$ = 30$^\circ$),
the high-velocity portion of ejecta ($v_{\rm ej}/v_{\rm imp}>$1.2) becomes bimodally distributed
and material is predominately jetted either immediately downrange (0$^\circ$) 
or at an angle of $\sim$ 40$^\circ$ (Figures \ref{fig:views}c and \ref{fig:hms}c). 
Note that ejection angles in iSALE-3D have some preferences, such as 45$^\circ$ and 135$^\circ$ (Figure \ref{fig:hms}a),
due to the use of cartesian coordinates.
While such a bimodal distribution is only observed for a 30$^\circ$ impact,
this behavior may be similar to the distal distribution of ejecta around elongated craters on Moon and Mars, 
which originate from oblique impacts \citep{Gault:1978aa,Schultz:1982aa}.
The volume of ejecta produced for a grazing impact ($\theta_{\rm imp}$ = 15 $^\circ$) 
scenario appears minimal (Figures \ref{fig:views}d and \ref{fig:hms}d), 
and high-velocity ejecta ($v_{\rm ej}/v_{\rm imp} > 0.5$) is only distributed downrange of the impact site.
We find almost all of high-velocity ejecta originate from the impactor.

\begin{figure}
\includegraphics[clip,width=0.5\textwidth]{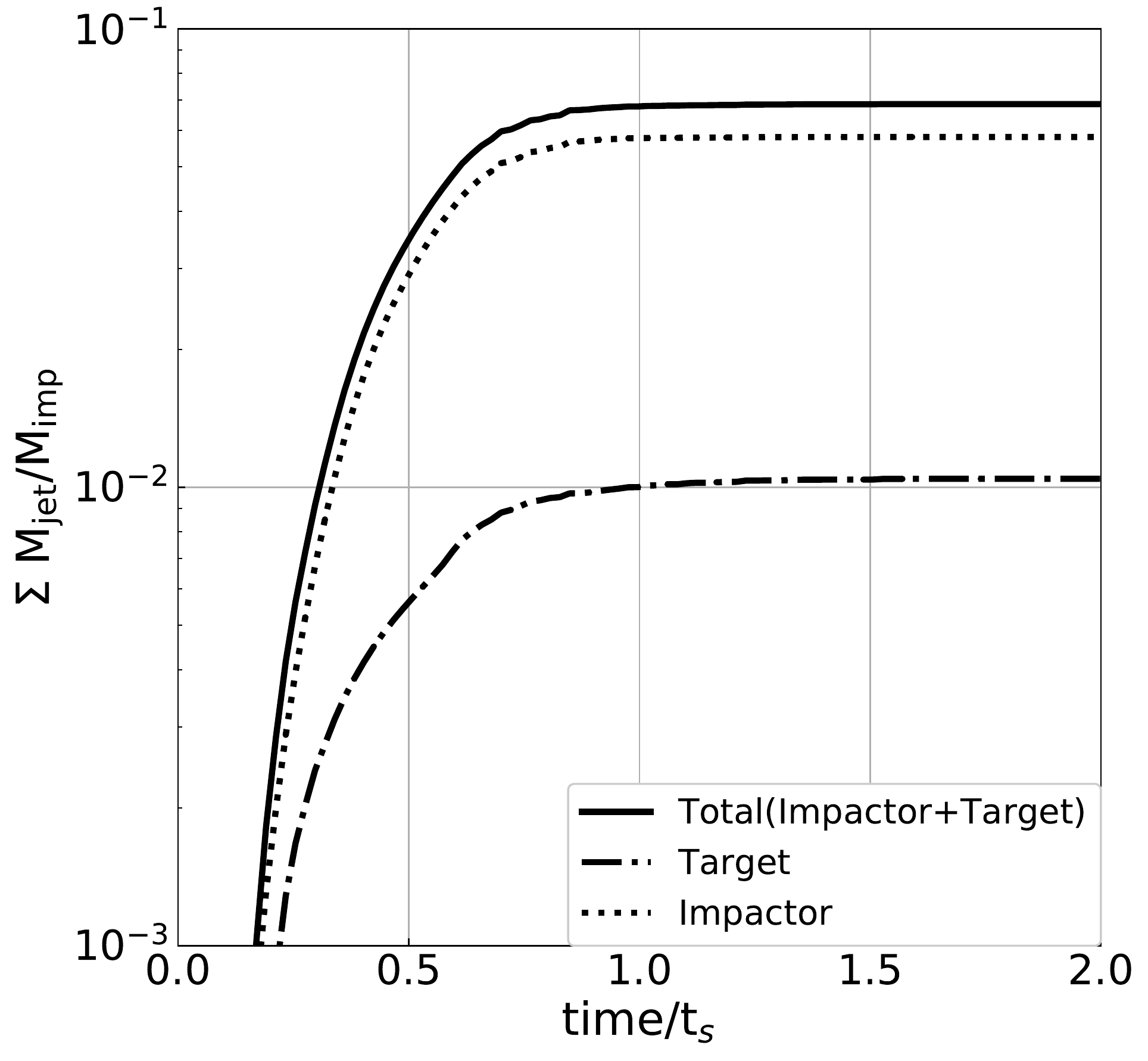}
\caption{
Evolution of cumulative jetted material mass over time for our fiducial model ($v_{\rm imp}$ = 3 km/s and $\theta_{\rm imp}$ = 45$^\circ$).
Each line represents jetted mass sourced from the target (dotted-dashed line), 
the impactor (dotted line), and the total amount of material ejected (solid line), respectively.
\label{fig:45tm}}
\end{figure}

\begin{figure}
\includegraphics[clip,width=0.5\textwidth]{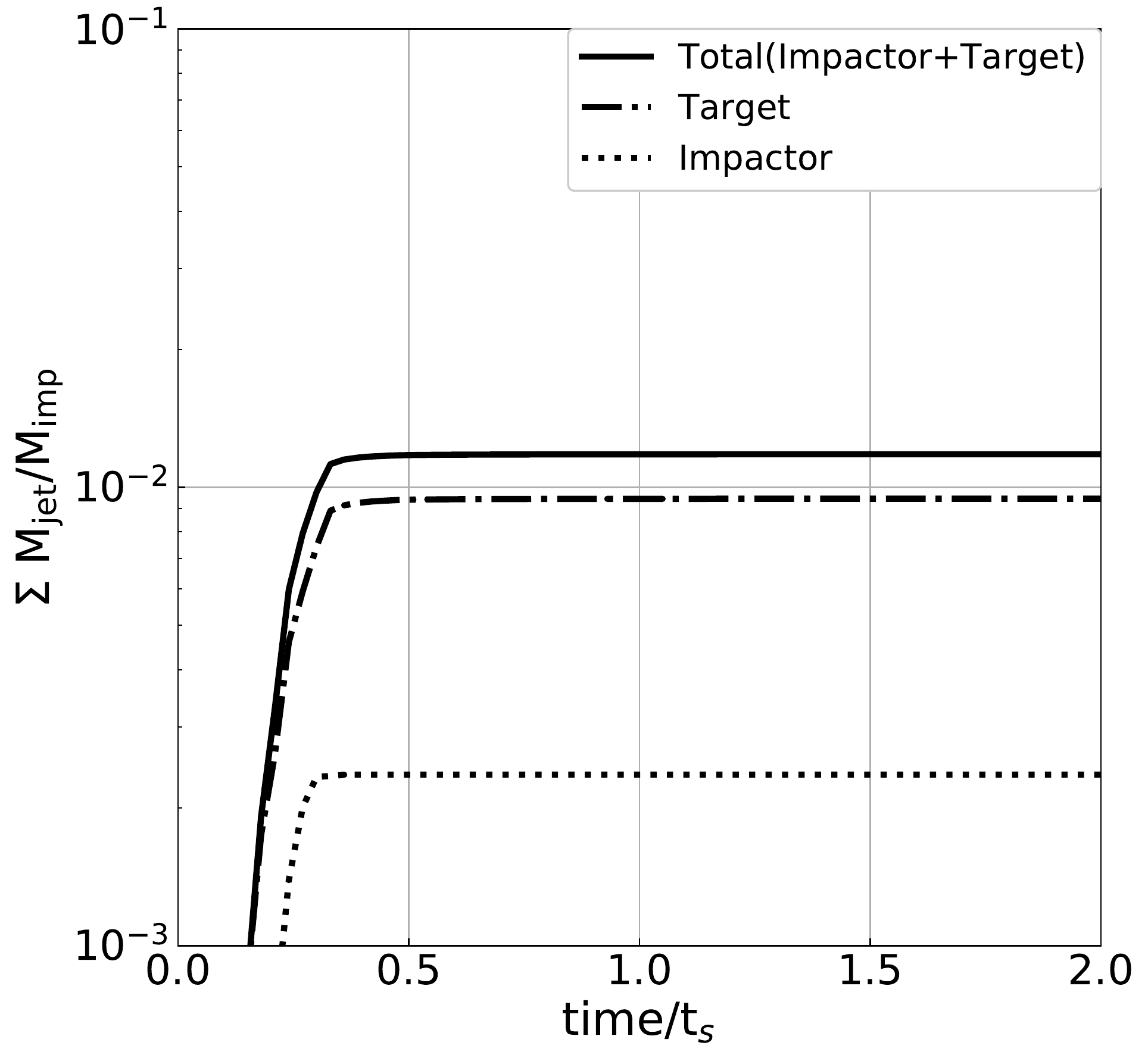}
\caption{Additional plots of jetted material mass over time as in Figure \ref{fig:45tm}, but for $\theta_{\rm imp}$ = 90$^\circ$.
\label{fig:90tm}}
\end{figure}

\begin{figure}
\includegraphics[clip,width=0.5\textwidth]{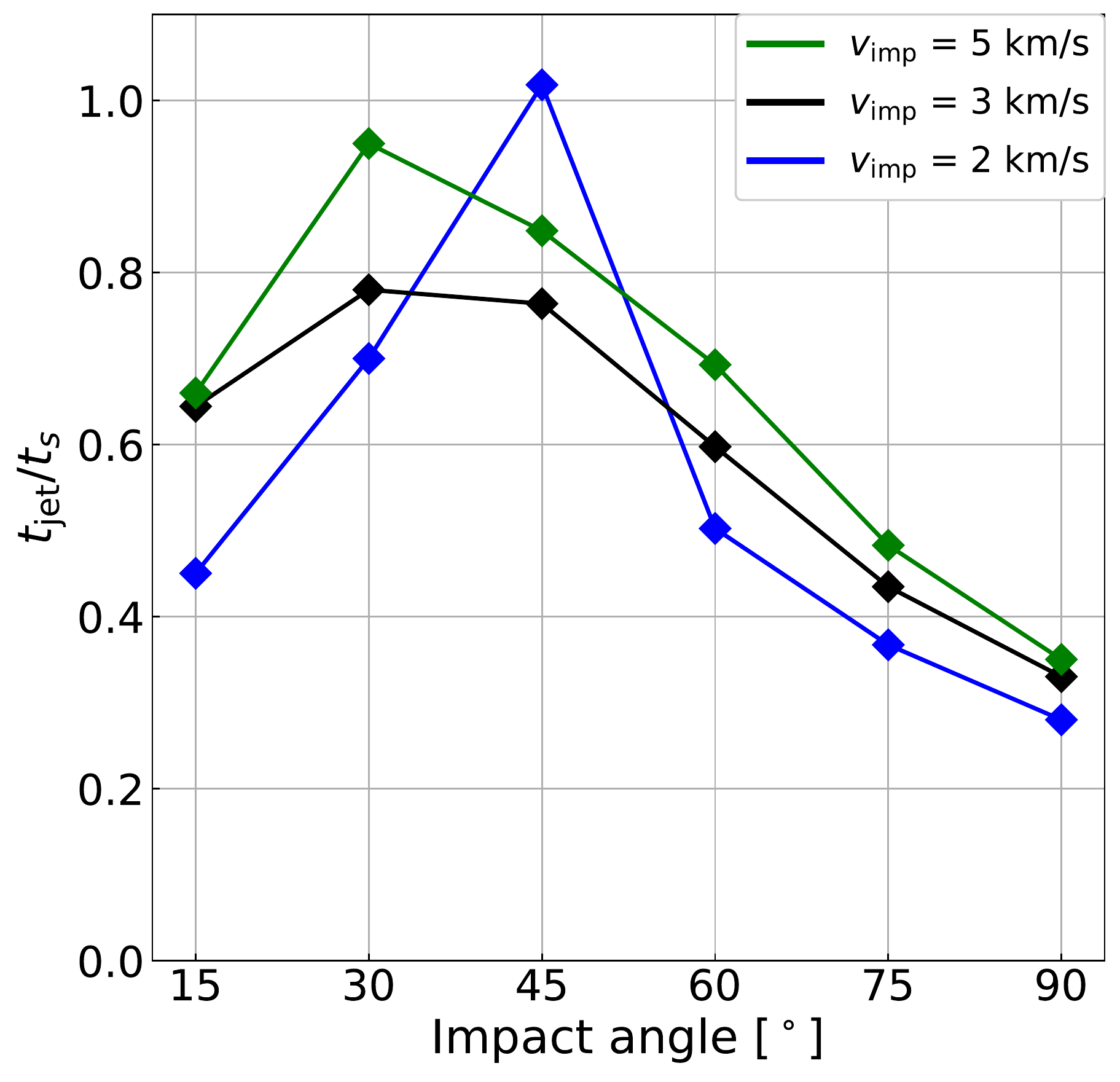}
\caption{
Time of jetting ($t_{\rm jet}$) as a function of impact angles. 
Color represent different impact velocities: 2 km/s (blue), 3 km/s (black), and 5 km/s (green).
\label{fig:tjet}}
\end{figure}

\subsection{Amount of jetted ejecta in oblique impacts} \label{sec:mjet}
We use Lagrangian tracer particles to quantify the amount of jetted ejecta for different impact angle and impact velocity scenarios. 
We define tracer particles as jetted ejecta when the tracer velocity exceeds the impact velocity, 
the vertical component of tracer velocity is positive, and the material is located above the surface of the target. 
Following the evolution of tracer particles that satisfy these conditions, our analysis indicates that 
as the impact progresses more material is ejected until jetting ceases. 
The earliest ejecta is the fastest while the latest is ejected just above the impact velocity.
To calculate for the cumulative jetted mass, we sum all tracer particles that fit this definition of jetted ejecta during impact.
Figure \ref{fig:45tm} illustrates the evolution of the cumulative jetted mass $M_{\rm jet}$ normalized by the impactor mass $M_{\rm imp}$ 
over time for our fiducial model ($v_{\rm imp}$ = 3 km/s and $\theta_{\rm imp}$ = 45$^\circ$).
The jetted material reaches 90\% of the total cumulative mass at $t$ = 0.76 $t_s$.
However, for a head-on impact, this peak occurs much earlier as $t$ = 0.33 $t_s$ (Figure \ref{fig:90tm});
this delay in jetting shows that impact jetting occurs over longer timescales for oblique impacts
(note $t_s = D_{\rm imp}/(v_{\rm imp} \sin(\theta_{\rm imp}))$).
Figure \ref{fig:tjet} represents the time of jetting $t_{\rm jet}$, when $\Sigma M_{\rm jet}$ reaches 90\% of its final value, as a function of impact angle. 
Even when scaled by the contact and compression timescale, the jetting timescale initially increases as impact angle decreases at a given impact velocity. 
This is likely due to the downrange momentum of the impactor and a favorable, less asymmetric, geometry for jetting. 
Simulations without material strength show this trend continues to $\theta_{\rm imp}$ = 30$^\circ$ for all impact velocities, 
demonstrating that material strength causes jetting to cease earlier in the case of $\theta_{\rm imp}$ = 30$^\circ$.
The jetting timescale for $\theta_{\rm imp}$ = 15$^\circ$ is shorter than other oblique impacts (Figure \ref{fig:tjet}).
For grazing impacts downward velocity is reduced before the contact and compression phase is over,
which sometimes results in decapitation of the impactor \citep{Schultz:1990ab,Davison:2011aa}. 
When this occurs, the impactor does not penetrate the target and jetting ceases earlier than in moderately oblique impacts.

\begin{figure}
\includegraphics[clip,width=\textwidth]{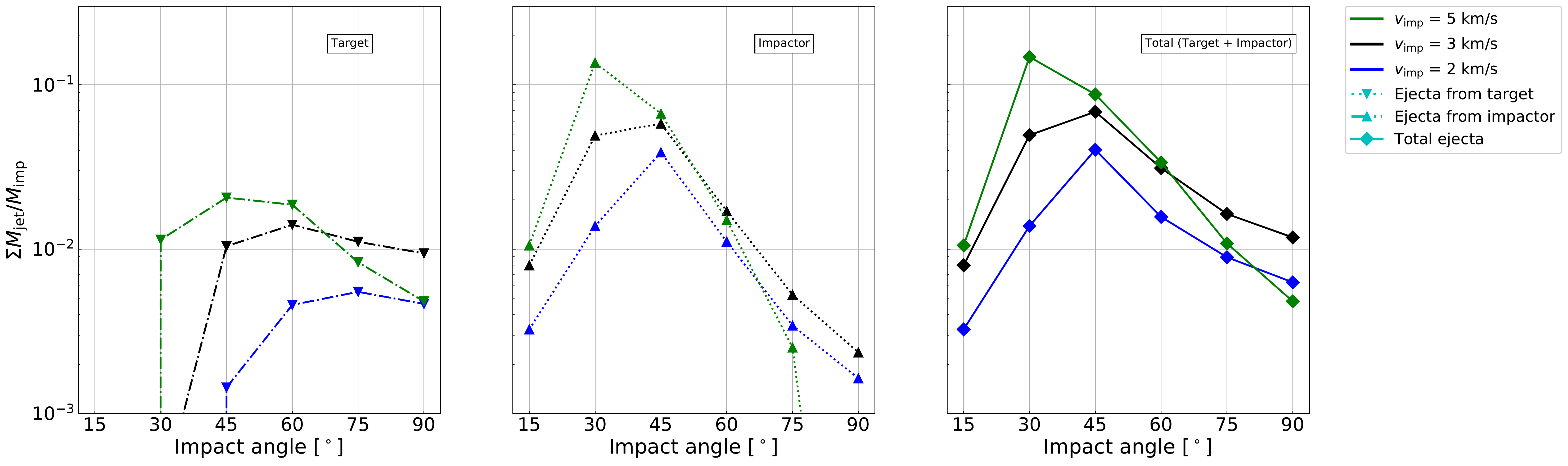}
\caption{
Mass of jetted material ($M_{\rm jet}$) as a function of impact angles. 
Color represent different impact velocities: 2 km/s (blue), 3 km/s (black), and 5 km/s (green).
Each panel depicts the total mass $M_{\rm jet}$ normalized by $M_{\rm imp}$ from the target (left), 
impactor (middle), and the total mass (right), respectively.
\label{fig:sum}}
\end{figure}

Figure \ref{fig:sum} summarizes the trends in cumulative jetted material mass $\Sigma M_{\rm jet}$ for all impact scenarios. 
Each panel represents $\Sigma M_{\rm jet}$ separated by material origin:
material originating from the target (dotted-dashed lines, left), impactor (dotted line, middle), 
and the total mass (solid line, right), respectively.
We first focus on the results of $v_{\rm imp}$ = 3 km/s (black lines), our fiducial velocity.
In this scenario, 
almost all oblique impacts ($\theta_{\rm imp}$ = 75$^\circ$, 60$^\circ$, 45$^\circ$, and 30$^\circ$) 
produce more jetted ejecta than the head-on impacts: 
1.6\%, 3.1\%, 6.8\%, and 4.9\% jetted mass normalized by impactor mass ($M_{\rm imp}$), respectively, 
in comparison to 1.1 \% for a vertical impact.
Jetted ejecta from our grazing impact ($\theta_{\rm imp}$ = 15$^\circ$) in this scenario is less than 1\% (Figures \ref{fig:views}d and \ref{fig:hms}d).
$\theta_{\rm imp}$ = 45$^\circ$ produces the most jetted mass ($\sim$ 7\%).

\begin{figure}
\includegraphics[clip,width=0.5\textwidth]{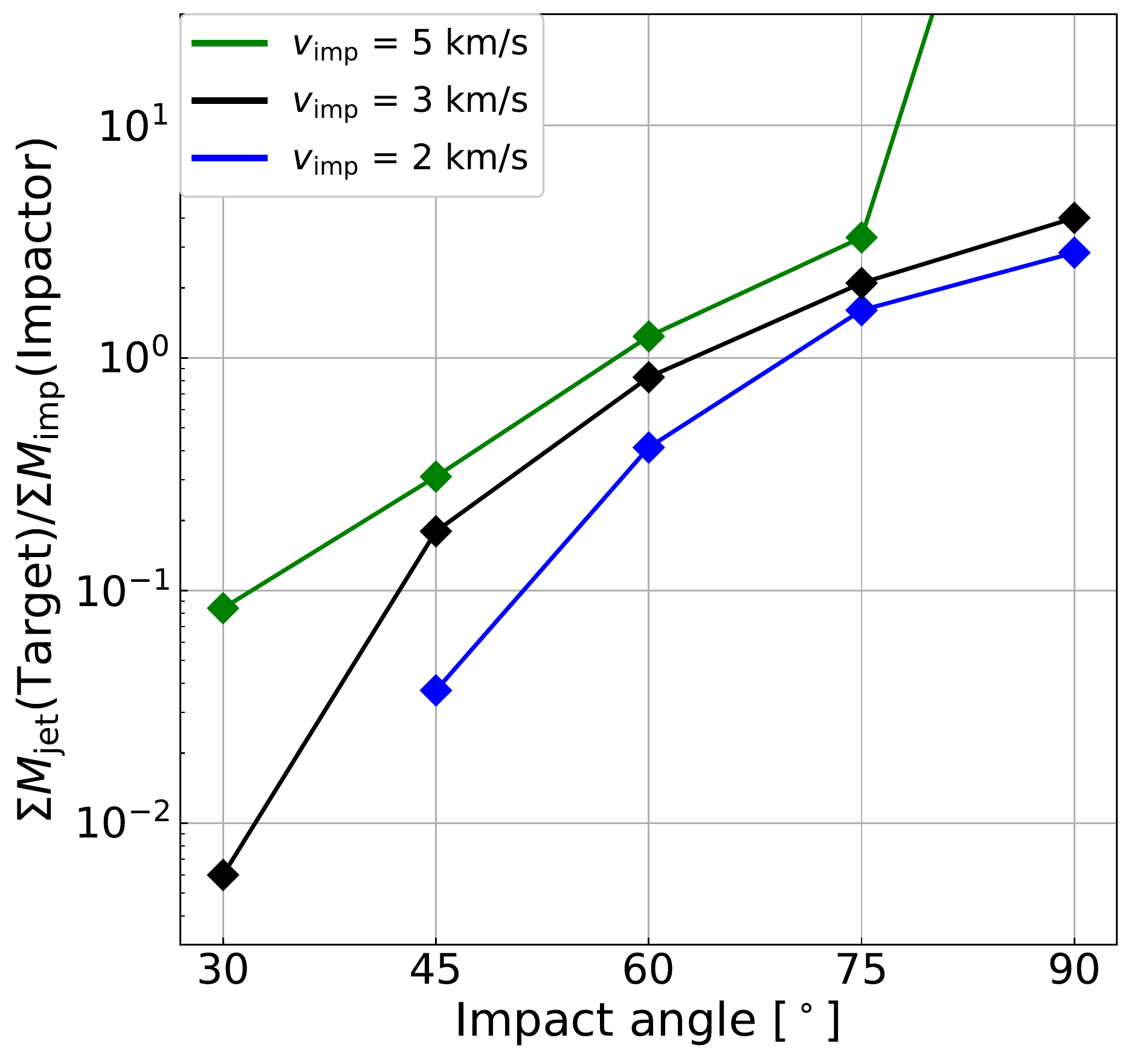}
\caption{Mass ratio of jetted material from the target to the impactor as a function of impact angles. 
Note that no symbols (e.g., $15^\circ$ cases) means no jetted material (see Figure \ref{fig:sum}).
\label{fig:sumratio}}
\end{figure}

The cumulative jetted mass in our fiducial velocity scenario ($v_{\rm imp}$ = 3 km/s) depends on impact angle, 
which also influences the source of jetted material (Figure \ref{fig:sum}).
As Figure \ref{fig:45tm} indicates, material from the impactor dominates the cumulative mass of jetted ejecta 
in our fiducial model ($v_{\rm imp}$ = 3 km/s and $\theta_{\rm imp}$ = 45$^\circ$). 
In contrast, jetted ejecta is dominated by target material for head-on impacts (Figure \ref{fig:90tm});
this distinction may result from the downrange momentum of the oblique impactor.
The fraction of impactor to the total jetted mass for our nominal impact velocity is 
84\% for a 45$^{\circ}$ impact and 20\% for a head-on impact. 
This trend is consistent with the results of laboratory experiments in previous work \citep{Schultz:1996aa,Sugita:1999aa}.
These experiments showed that oblique impacts tend to produce more jetted material from the impactor. 
Our results are consistent with this trend: 
Figure \ref{fig:sumratio} clearly illustrates that the ratio of jetted material from the target to the impactor decreases as the impact angle gets shallower.

Our results for a lower impact velocity ($v_{\rm imp}$ = 2 km/s, blue lines in Figure \ref{fig:sum}) indicate that 
$\Sigma M_{\rm jet}$ also depends on impact velocity.
In this scenario, the jetted mass produced in head-on impacts is 0.6\% of the impactor mass, almost half of that produced for $v_{\rm imp}$  = 3 km/s.
Oblique impacts also produce less jetted ejecta at a lower velocity; however,
we see similar trends with impact angle and 
$\theta_{\rm imp}$ = 45$^\circ$ still produces the largest mass of jetted material (4.0\% of the impactor mass).
Since the jetting timescale of 45$^\circ$ impact is much longer than other impacts with $v_{\rm imp}$ = 2 km/s (Figure \ref{fig:tjet}), 
jetting in $\theta_{\rm imp} = 45 ^\circ$ continues longer then produces more jetted mass.
A lower jetted mass at a lower (2 km/s) impact velocity 
contradicts previous simulations using a strengthless material \citep{Johnson:2014aa}, 
likely because our incorporation of material strength limits the amount of jetted ejecta produced by low velocity impacts. 
This behavior is supported by previous modeling work \citep[][see also \ref{sec:app}]{Johnson:2015aa,Wakita:2017aa}.

When $v_{\rm imp}$ is increased ($v_{\rm imp}$ = 5 km/s, green lines in Figure \ref{fig:sum}), 
the resulting behavior of the jetted ejecta can be explained by two competing effects. 
As impact velocity increases, jetting initiates later \citep[e.g.,][]{Sugita:1999aa,Johnson:2014aa}.
For $\theta_{\rm imp}$ = 90$^\circ$ and 75$^\circ$ this effect reduces the amount of jetted material as compared to our simulations with $v_{\rm imp}$ = 3 km/s.
While the time when jetting ceases is almost the same for $\theta_{\rm imp}$ = 90$^\circ$ and 75$^\circ$ regardless of impact velocity, 
jetting ceases earlier for lower-velocity impacts occurring at lower impact angles (Figure \ref{fig:tjet}).
Thus, despite increasing impact velocity, jetted mass for  $\theta_{\rm imp}$ = 45$^\circ$ and 60$^\circ$ are similar to the fiducial case. 
As impact angle is decreased ($\theta_{\rm imp}$ = 30$^\circ$), 
the time of jetting increases and more jetted mass is produced than our fiducial case (14\% compared to 6.8\%), 
which is the highest value among impact with $v_{\rm imp}$ = 5 km/s. 
Simulations without material strength show that jetted mass decreases as impact velocity increases regardless of impact angle, 
which is consistent with previous work \citep{Johnson:2014aa}. 
Thus, material strength has an important effect on the jetting efficiency during oblique impacts in the velocity range explored.

\section{Discussion} \label{sec:dis}

\begin{figure}
\includegraphics[clip,width=\textwidth]{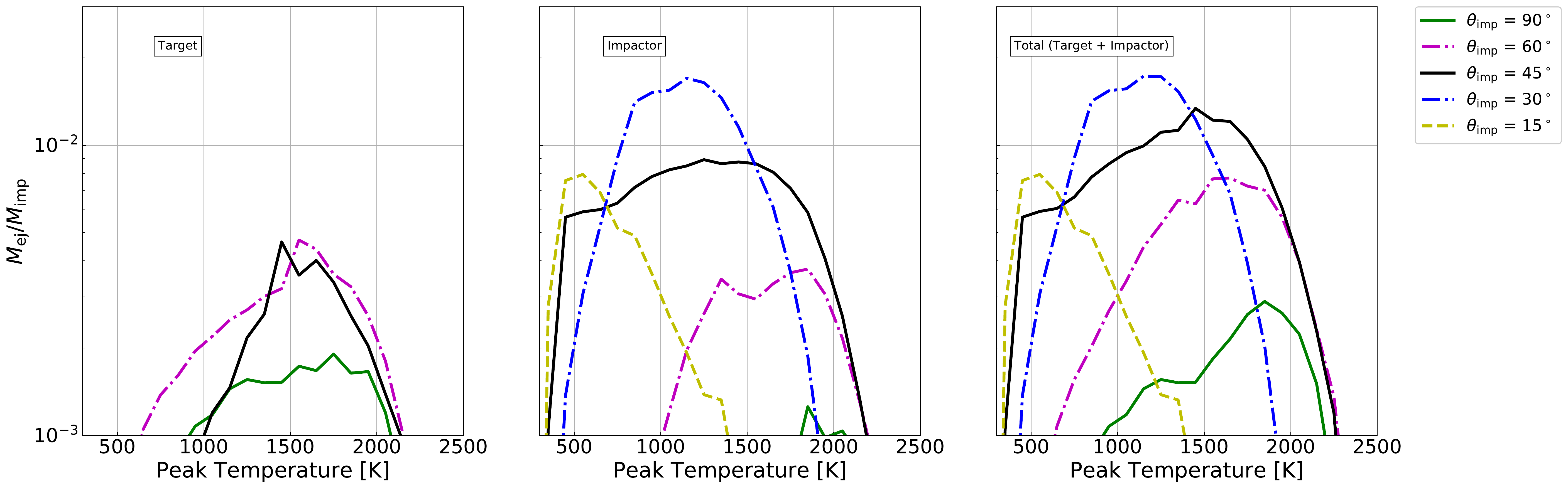}
\caption{
Histogram of peak temperature of ejecta shown in Figures \ref{fig:45}(c) and \ref{fig:views} 
($v_{\rm ej}/v_{\rm imp} > 0.5$ at $t_{\rm jet}$ for  $v_{\rm imp}$ = 3 km/s).
Bin size of peak temperature is 100 K.
Mass of ejecta ($M_{\rm ej}$) is normalized by mass of impactor $M_{\rm imp}$. 
\label{fig:hist_temp}}
\end{figure}

\begin{figure}
\includegraphics[clip,width=\textwidth]{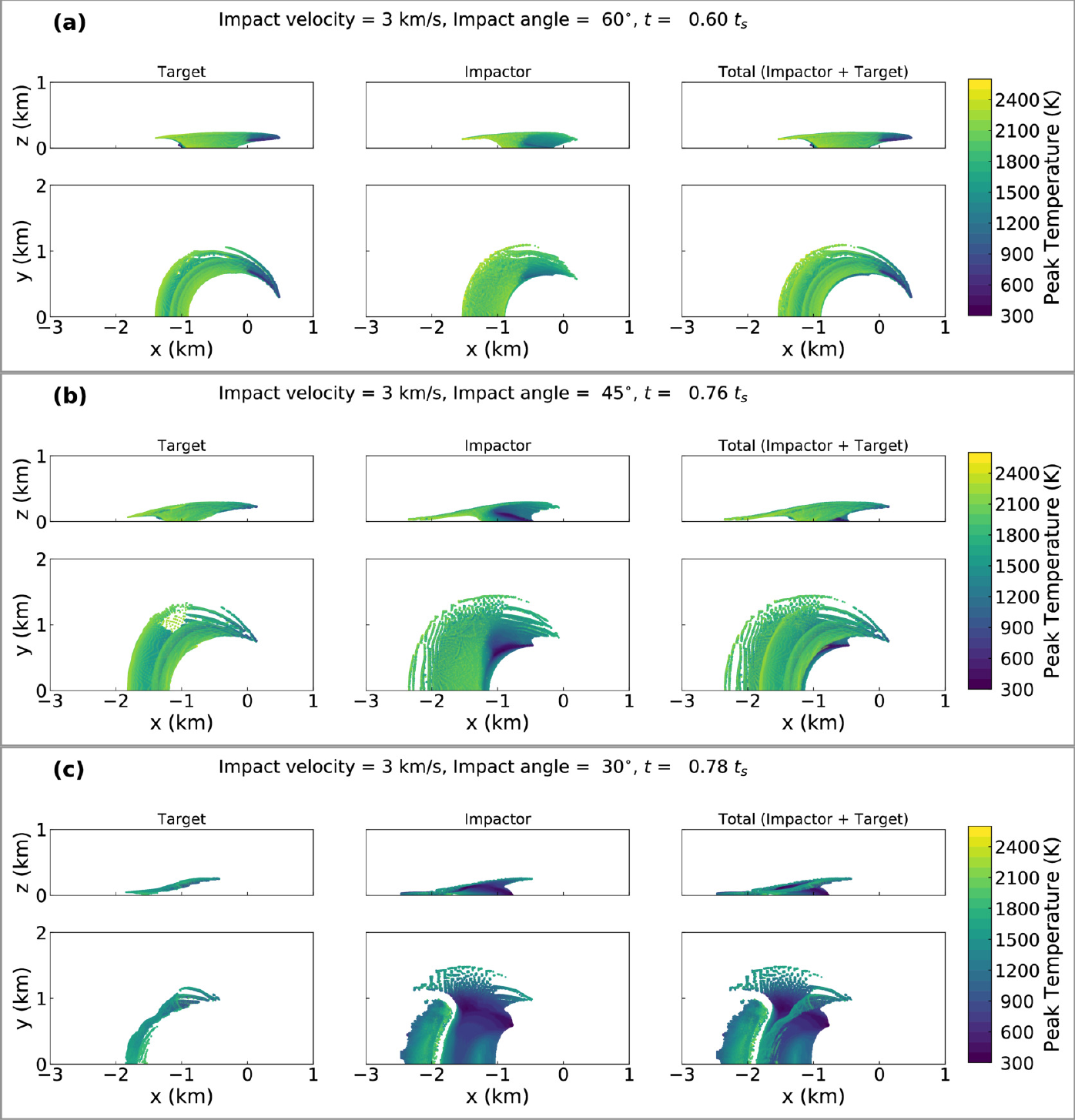}
\caption{
Behavior of ejected material for 
$\theta_{\rm imp}$ = 60$^\circ$, 45$^\circ$, and 30$^\circ$ (top to bottom).
Same viewing geometries and time as Figures \ref{fig:45}(c) and \ref{fig:views}, but color represents peak temperature in Kelvin.
\label{fig:temp}}
\end{figure}

\begin{figure}
\includegraphics[clip,width=\textwidth]{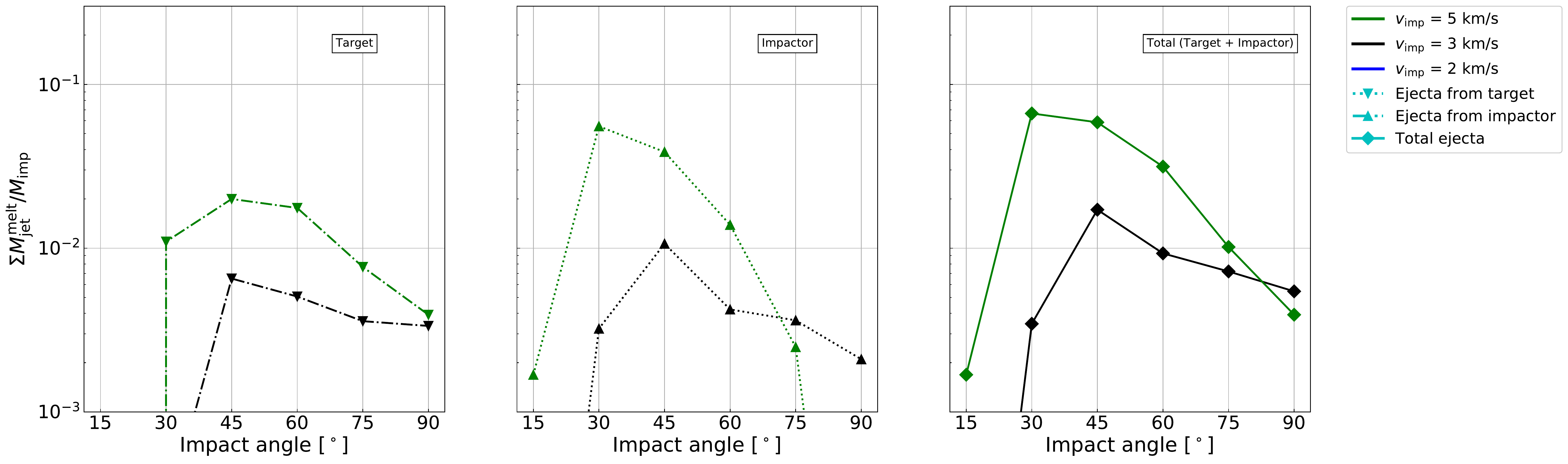}
\caption{Same as Figure \ref{fig:sum}, but for $\Sigma M_{\rm jet}^{\rm melt}$. 
Note that $\Sigma M_{\rm jet}^{\rm melt}$ of 2 km/s is less than $10^{-3}$ and cannot be seen here.
\label{fig:melt}}
\end{figure}

\begin{figure}
\includegraphics[clip,width=\textwidth]{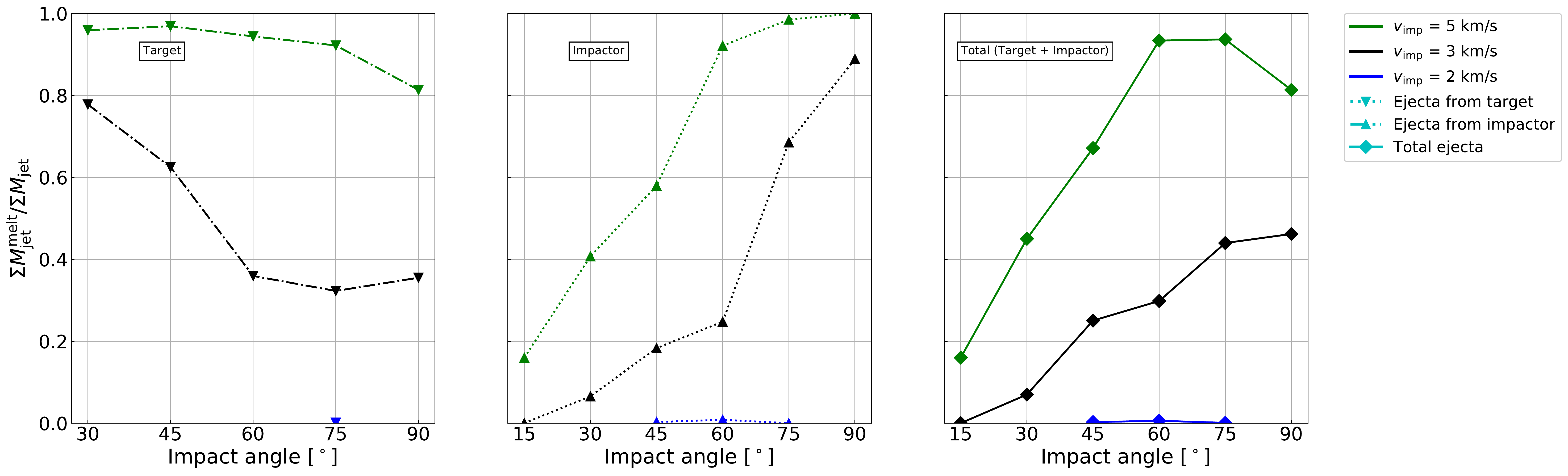}
\caption{Same as Figure \ref{fig:sum}, but for $\Sigma M_{\rm jet}^{\rm melt}/\Sigma M_{\rm jet}$. 
Note that no symbols for 2km/s cases means no jetted melt.
\label{fig:meltfrac}}
\end{figure}

Our results show that the efficiency and dynamics of the jetting process are sensitive to impact angle and impact velocity.
In general, oblique impacts tend to produce less shock-heated material than vertical impacts \citep{Pierazzo:2000aa, Davison:2014aa}. 
However, this trend does not hold when considering only jetted material.
Previous work suggests that the fraction of jetted material that reaches high temperatures (i.e., may experience melting and vaporization)
will also increase as the impact becomes more oblique \citep{Melosh:1986aa,Vickery:1993aa}.
Figure \ref{fig:hist_temp} represents the histogram of peak temperature of ejecta shown in Figures \ref{fig:45}(c) and \ref{fig:views} 
($v_{\rm ej}/v_{\rm imp} > 0.5$ at $t_{\rm jet}$ for  $v_{\rm imp}$ = 3 km/s).
The spatial distribution of fast ejecta colored according to peak temperature (Figure \ref{fig:temp}) 
demonstrates that the fastest ejecta, preferentially directed downrange, also reaches the highest temperatures.
The peak of each histogram (Figure \ref{fig:hist_temp}) shifts to higher temperature as the impact angle increases.
This seems broadly consistent with laboratory experiments, which showed that 
the temperature of jetted material increases with impact angles (i.e., vertical impacts produce the hottest jet) \citep{Sugita:1998aa,Sugita:1999aa}.
To track the evolutional history of their temperature (i.e., they have experienced melting),
we expand the analyze and quantify the potential amount of jetted melt from our simulations.
We define jetted melt as jetted ejecta that exceeds the solidus temperature of dunite (1373K) 
after the pressure has dropped below
1 bar ($10^5$Pa).
When jetted ejecta have reached high pressure, we track them until their pressure becomes lower than this threshold, 
then examine their temperature as a post-shock temperature to compare with the solidus temperature.
Figure \ref{fig:melt} shows the total amount of jetted melt $M_{\rm jet}^{\rm melt}$ normalized by $M_{\rm imp}$ as a function of the impact angle.
Although the melted ejecta of most cases for $v_{\rm imp}$ = 3 km/s fail to reach 1\% of $M_{\rm imp}$, 
likely due to the lower resolution of the model run,
melted material for $v_{\rm imp}$ = 5 km/s exceeds 1\% of $M_{\rm imp}$ (green lines in Figure \ref{fig:melt}),
which we consider to be a significant melt mass.
We find that melt fraction in jetted ejecta depends on the impact angle, and that melt fraction decreases as impact angles get shallower (Figure \ref{fig:meltfrac}),
which directly conflicts with previous work based on thin plate theory \citep{Vickery:1993aa} 
suggesting that jets produced by grazing impacts have a higher melt content than the jet produced by vertical impacts. 
However, because the jetted mass produced by moderately oblique impacts is larger than that of vertical impacts (Figure \ref{fig:sum}), 
so is the amount of melt. 
Comparison to high resolution axisymmetric simulations shows that 
jetted melt in our $v_{\rm imp}$ = 5 km/s is underestimated by a factor of 1.2 similar to the resolution of jetted mass. 
The same comparison for $v_{\rm imp}$ = 3 km/s, however, shows the jetted melt is underestimated by a factor of 2.1 (\ref{sec:app}). 
This result is not surprising as previous simulations of \citet{Johnson:2015aa} show that the early fastest-moving portion of the jet, 
which is more difficult to resolve, is dominated by melt. 


If jetted ejecta are melt-dominated (Figures \ref{fig:melt} and \ref{fig:meltfrac}), 
this may increase the amount of chondrules produced during planetary accretion \citep[e.g.,][]{Kieffer:1975aa,Johnson:2015aa,Wakita:2017aa}. 
Chondrules are millimeter-sized spherical materials found in primitive meteorites \citep[e.g.,][]{Scott:2014aa}. 
While the processes that produce chondrules are still debated,
impact jetting during accretionary impacts is one proposed formation mechanism.
In this scenario, melt jetted during a relatively low velocity collision may break up and cool, ultimately becoming chondrules \citep{Johnson:2015aa,Johnson:2018aa}.
The amount of jetted melt from our three-dimensional vertical impact simulations is smaller 
than previous two-dimensional calculations, likely due to our lower resolution (see also \ref{sec:app}).
Our work, however, indicates that oblique impacts produce more jetted melt than vertical impacts (Figure \ref{fig:melt}).
Given that oblique impacts occur more frequently \citep{Shoemaker:1962aa},
they may have enhanced efficiency in chondrule formation in the early Solar System.

Our results also show that the amount of jetted ejecta originating from the impactor increases as impact angles become shallower (Figure \ref{fig:sum});
this trend suggests that the amount of melted material sufficient to form chondrules 
will have a greater component of impactor material than head-on impacts at the same impact velocity.
Chondrule precursor material is thought to be pristine undifferentiated material \citep{Taylor:1983aa}.
While smaller impactors tend to remain undifferentiated, the larger target might be differentiated \citep{Gail:2014aa, Lichtenberg:2018aa, Wakita:2018aa}, 
though such larger bodies might have an undifferentiated surface layer \citep[e.g.,][]{Weiss:2013aa}. 
Our findings that jetted material produced during oblique impacts is dominated by impactor material
may suggest that an undifferentiated impactor can produce chondrules with primitive compositions even if the target is completely differentiated. 

Our results may also have implications for the origin of unique distal ejecta patterns observed around craters produced by oblique impacts.
Distal distribution of ejecta, known as butterfly patterns, are found around elongated craters on Moon and Mars \citep{Gault:1978aa,Schultz:1982aa}.
As Figures \ref{fig:views}c and \ref{fig:hms}c indicate, there is 
a discontinuous distribution of jetted material for $\theta_{\rm imp}$ = 30$^\circ$ 
(the gap around $\sim 30^\circ$ of (azimuthal) ejection angle in Fig. \ref{fig:hms}c, 
which corresponds to around (x, y) = (-2 km, 1 km) in Fig. \ref{fig:views}c). 
Slower ejecta released at later stages would be the primary source of an asymmetric ejecta pattern, rather than the earlier jetted ejecta.
It is unclear from our work if the asymmetric pattern for fastest ejecta would affect the distribution of later slower ejecta. 
Additionally, we note the impact velocities explored here are significantly lower than typical impact velocities on the Moon 
and other terrestrial planets so these comparisons may be limited.
s
Our results may also explain differences in reflectance in distal ejecta on the Moon 
as a difference in material composition between target and impactor \citep{Speyerer:2016aa}.
While the impact velocity on the Moon ranges over 6 km/s to 40 km/s \citep{Chyba:1991aa,Yue:2013aa}, 
the escape velocity of the Moon is 2.4 km/s. 
Thus, jetted ejecta as defined here will exceed the escape velocity of the Moon.
As previously noted, the fastest portions of the jet eject at very low angles and graze the surface.
Considering a realistic target with topography and slopes some of this material will impact the surface and be deposited even though it is ejected above escape velocity.
Our results for grazing impacts ($\theta_{\rm imp}$ = 15$^\circ$) record a deficit in ejecta velocity, 
(around $0.5v_{\rm imp}$)
resulting from the difference in ejecta velocity between ejecta originating from target and impactor (Figures \ref{fig:views}d and \ref{fig:hms}d).
If the composition of the impactor differs from that of the target, 
ejecta of oblique impacts originating from the impactor might produce distal ejecta  
with different reflectance characteristics than proximal ejecta, 
which may originate from the target \citep{Speyerer:2016aa}.

Lastly, we assess the influence of resolution on our results by comparing 
our three dimensional simulations with two-dimensional calculations. 
Our three-dimensional results are in good agreement 
with two-dimensional simulations at the same resolution of 5 m or 100 cells per projectile radius (CPPR); however,
higher resolution is required to produce nearly same amount of jetted material as fully resolved two-dimensional calculations (see \ref{sec:app}):
The jetted mass in our simulations with 100 CPPR are 60\%-70\% less than that produced in our 1000 CPPR resolution simulations.
This finding is consistent with previous work reporting that resolution $>$400 CPPR is necessary 
to resolve high-velocity jetted material based on two-dimensional simulations \citep{Johnson:2014aa,Wakita:2017aa,Kurosawa:2018ab}.
However, because the cumulative jetted mass produced by oblique impacts is more than twice the amount produced in head-on impacts, 
our simulations should adequately resolve the influence of impact angle on the mass and the distribution of impactor and target materials in jetted ejecta.
It is difficult and computationally expensive to run the three-dimensional simulations in high resolution, however,
limiting the high-resolution zone as done in this paper will help to tackle this.
Our results provide an important step toward understanding jetting during oblique impacts.

\section{Conclusions} \label{sec:con}
Our simulations of oblique impacts illustrate that the amount of jetted materials and their origin strongly depends on impact angle.
For an impact velocity of 3 km/s, 
a 45$^\circ$ impact produces six times the mass of jetted materials relative to a vertical impact.
The source of jetted material varies; for $<$ 45$^\circ$ impacts, most jetted material originates from the impactor;
however, target material begins to be the primary source of jetted ejecta at shallower impact angles (75$^\circ$ and 90$^\circ$).
While jetted ejecta from oblique impacts is distributed downrange of the impactor (Figures \ref{fig:45} and \ref{fig:views}),
grazing impacts (impact angles $<$ 30$^\circ$) have a more discontinuous distribution.
Such distribution by grazing impacts may help explain the distal ejecta patterns observed around elongated craters. 
Additional higher resolution simulations of oblique impacts can provide more accurate estimates of both the amount of jetted ejecta 
and its melt fraction, a potential source for chondrules.

\section*{Acknowledgments}
We gratefully acknowledge the developers of iSALE-3D, including Gareth Collins, Kai W{\"u}nnemann, Dirk Elbeshausen, Boris Ivanov and Jay Melosh.
Numerical computations were carried out on the PC cluster and the analysis servers at the Center for Computational Astrophysics, National Astronomical Observatory of Japan.
This work was supported by grant 80NSSC20K0422 from the NASA Emerging Worlds program.
T. M. D. was funded by STFC Grant ST/S00615/1.
We thank anonymous referees whose comments helped to improve this manuscript.

\appendix
\setcounter{figure}{0}
\setcounter{table}{0}
\section{Head-on impacts in two-dimensional calculation} \label{sec:app}

\begin{figure}
\includegraphics[clip,width=0.5\textwidth]{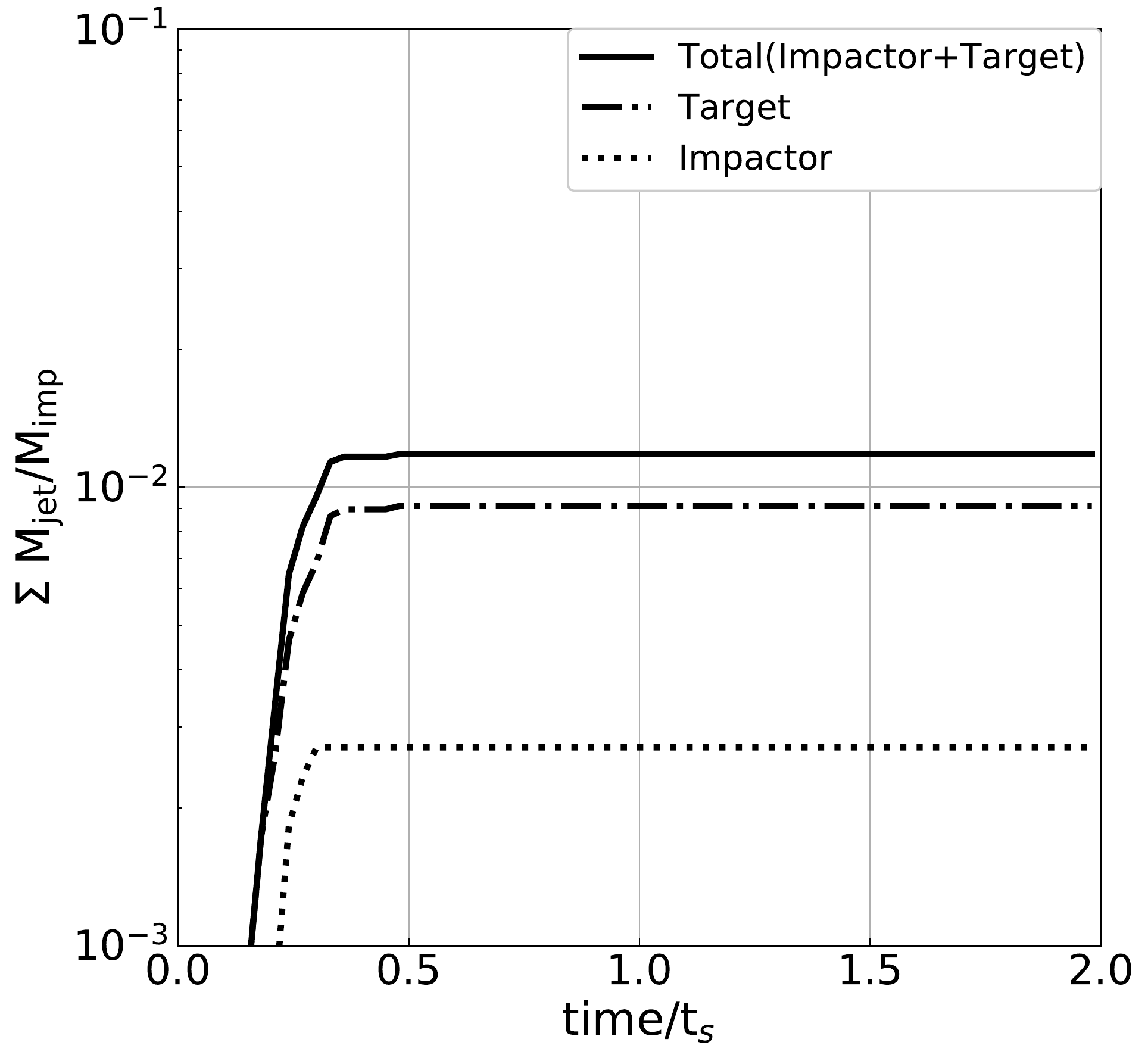}
\caption{Same evolution of jetted material over time as in Figure \ref{fig:90tm}, but using iSALE-2D (CPPR=100).
\label{fig:90tm2d}}
\end{figure}

\begin{figure}
\includegraphics[clip,width=\textwidth]{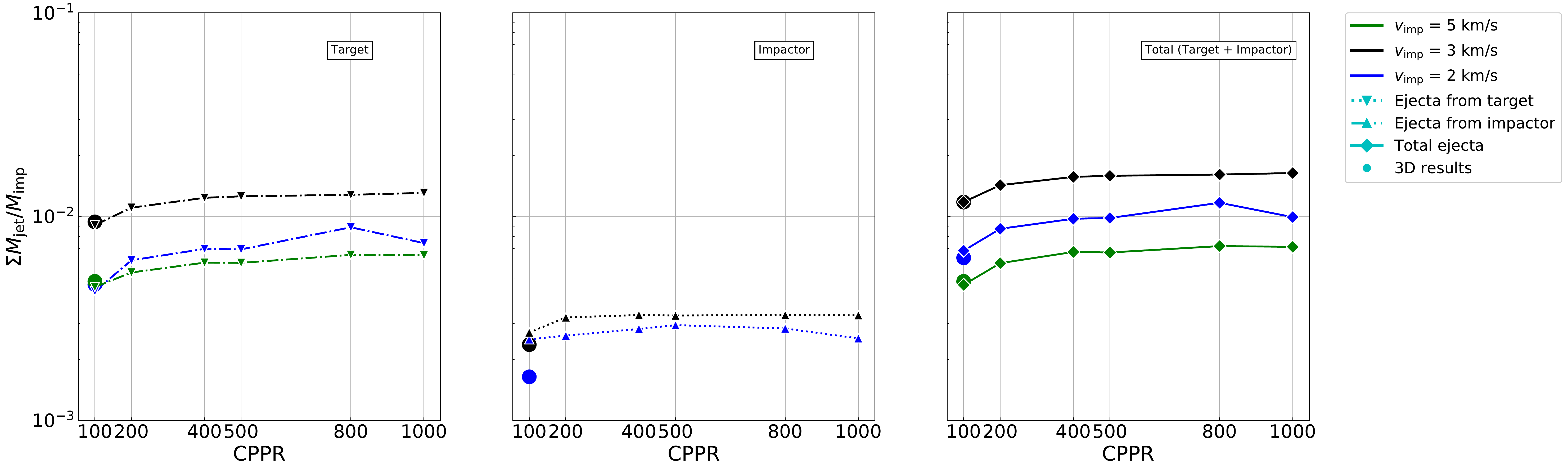}
\caption{Cumulative mass of jetted material ($\Sigma M_{\rm jet}$) as a function of CPPR for the head-on impacts. 
Color represents different impact velocities: 2 km/s (blue), 3 km/s (black), and 5 km/s (green).
The panels depict the total mass $M_{\rm jet}$ normalized by $M_{\rm imp}$ 
from the target (left), the impactor (middle), and the total mass (right), respectively.
The head-on impact results from iSALE-3D are plotted as circles.
Note that jetted ejecta from the impactor in the case of 5 km/s is less than $10^{-3}$ (middle panel).
\label{fig:cppr}}
\end{figure}

Here, we compare the results for the mass of jetted material produced in head-on impacts for iSALE-3D
to those using the iSALE-2D shock physics code.
iSALE-2D is based on the SALE hydrocode \citep[][version iSALE-Dellen]{Amsden:1980aa,Wunnemann:2006aa,Collins:2016aa}.
The highest resolution we used in iSALE-3D was 5m, with 100 cells per projectile radius (CPPR). 
For our iSALE-2D runs we include the entire spherical impactor in the high resolution zone,
instead of the lower part used in iSALE-3D (see Section \ref{sec:met} and Figure \ref{fig:45totalv}).
The comparison below demonstrates that the smaller high resolution zone used in our 3D simulations does not affect jetted mass.
Figure \ref{fig:90tm2d} shows the evolution of cumulative jetted mass produced from head-on impacts using iSALE-2D with 100 CPPR and $v_{\rm imp}$ = 3 km/s.
Although the details differ between Figures \ref{fig:90tm} (iSALE-3D) and \ref{fig:90tm2d} (iSALE-2D), 
the cumulative jetted mass is 1.1 \% of impactor mass for both scenarios.
Our results for $\Sigma M_{\rm jet}$ produced from head-on impacts in iSALE-2D as a function of resolution are are summarized in Figure \ref{fig:cppr}.
Our results for 100 CPPR from iSALE-2D and iSALE-3D with $v_{\rm imp}$ = 2 km/s and 5 km/s give similar results
($\Sigma M_{\rm jet}/M_{\rm imp}$ is 0.68 \% and 0.46 \% for iSALE-2D, and 0.62 \% and 0.48 \% for iSALE-3D, respectively).

\begin{figure}
\includegraphics[clip,width=\textwidth]{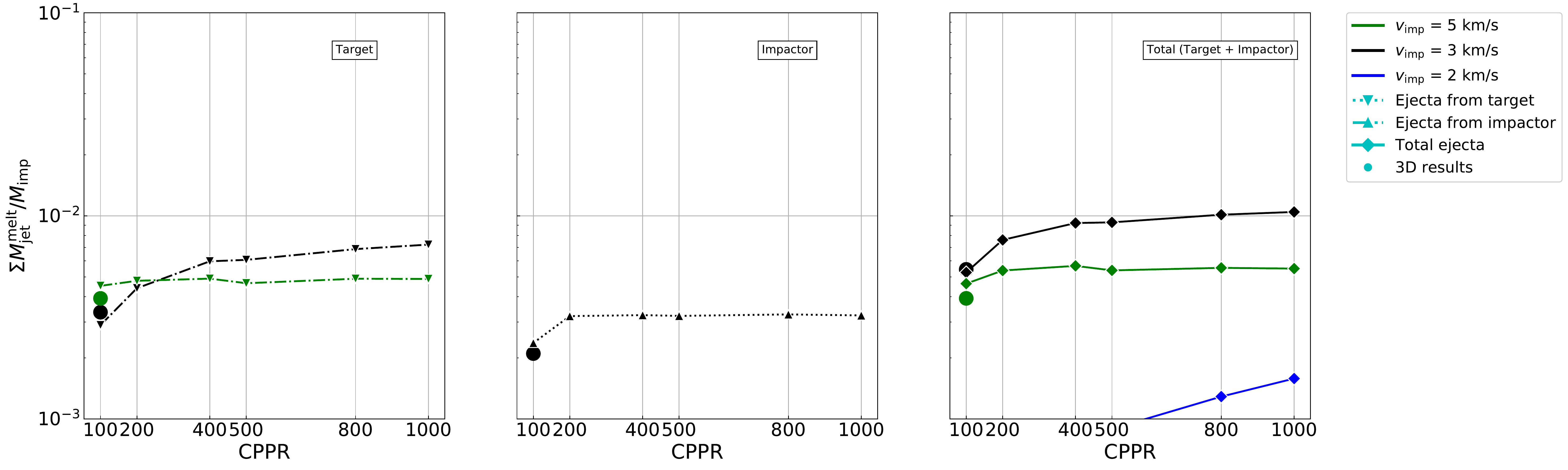}
\caption{Same as Figure \ref{fig:cppr}, but for $\Sigma M_{\rm jet}^{\rm melt}$. 
Note that jetted melt from the target at 2 km/s and the impactor at 2 km/s and 5 km/s are less than $10^{-3}$. 
\label{fig:cpprmelt}}
\end{figure}

Our results show that our simulations using iSALE-2D are well converged at 400 CPPR.
The jetted mass fraction compared to that produced from 1000 CPPR  are shown in Table \ref{tab:cppr}.
Taking our 1000 CPPR results as fully converged, 
our 3D head-on impact results with 100 CPPR underestimate the jetted mass by a factor of 1.4 to 1.6, depending on impact velocity. 
In simulations with 100 CPPR, there is little difference between the amount of jetted melt produced in iSALE-2D and iSALE-3D, 
but they underestimate jetted melt relative to 1000 CPPR simulations by a factor of 2.1 at 3 km/s and 1.2 at 5 km/s (Figure \ref{fig:cpprmelt}).
Note that jetted melt at 2 km/s with 800 and 1000 CPPR exceeds 0.1\% of $M_{\rm imp}$.
It remains unclear whether oblique impact simulations with higher resolution would produce more jetted melt than simulations at 100 CPPR.
As Figure \ref{fig:sum} illustrates, oblique impacts generate more jetted material than head-on impacts.
Therefore, our setting of 100 CPPR may be sufficient to characterize the jetted mass produced by oblique impacts, though not vertical impacts.
Regardless, we expect the trends in jetted mass as a function of impact angle and velocity for higher resolution simulations to be similar to those presented here.

\begin{threeparttable}
\caption{The jetted mass fraction ratio to that from CPPR=1000 \tnote{*}}
\begin{tabular}{lccccccl}
\hline
\hline
$v_{\rm imp}$ & CPPR  = 100 (3D) & 100 (2D) & 200 & 400 & 500 & 800 & 1000 \\
\hline
2 km/s & 0.63 & 0.68 & 0.87 & 0.98 & 0.99 & 1.1 & 1.0 (9.9$\times 10^{-3}$) \\
3 km/s & 0.72 & 0.72 & 0.87 & 0.95 & 0.96 & 0.98 & 1.0 (1.6$\times 10^{-2}$) \\
5 km/s & 0.67 & 0.65 & 0.83 & 0.94 & 0.93 & 1.0 & 1.0 (7.1$\times 10^{-3}$) \\
\hline
\end{tabular}
\begin{tablenotes}
\item[*]These values are respected to the jetted mass fraction of CPPR=1000 (see Figure \ref{fig:cppr}).
Absolute values of CPPR=1000 are shown in bracket.
\end{tablenotes}
\label{tab:cppr}
\end{threeparttable}

\end{document}